\documentclass{amsart}
\usepackage[colorlinks=true,pdftex]{hyperref}
\usepackage[latin1]{inputenc}
\usepackage[T1]{fontenc}
\usepackage{mathrsfs}
\usepackage{mathpazo}
\usepackage[abbrev]{amsrefs}

\newcommand{\hook}{
\mathbin{\mbox{\vrule width5pt height0.2pt\vrule height5pt width0.2pt}
}}

\newcommand{\cA}{\mathscr{A}}
\newcommand{\cC}{\mathscr{C}}
\newcommand{\cG}{\mathscr{G}}
\newcommand{\cL}{\mathscr{L}}
\newcommand{\cO}{\mathscr{O}}
\newcommand{\cP}{\mathscr{P}}

\newcommand{\RR}{\mathbb{R}}

\newcommand{\dd}{\mathrm{d}}
\newcommand{\del}{\partial}

\newcommand{\Lie}{\mathrm{L}}

\begin{document}

\title{On the applicability of constrained symplectic integrators in
  general relativity } 
 
\author{Jörg Frauendiener}
\address{Department of Mathematics and Statistics, University of
  Otago, P.O. Box 56, Dunedin 9010, New Zealand} 
\address{Centre of Mathematics for
  Applications, University of Oslo, P.O. Box 1053 Blindern, NO-0316
  Oslo, Norway}
\email{joergf@maths.otago.ac.nz}
\thanks{Discussions with C.~Lubich and R.~Richter on the numerical
  implementation of the Einstein equations using symplectic
  integrators are gratefully acknowledged.}
\date{\today}

\begin{abstract}  
  The purpose of this note is to point out that a naive application of
  symplectic integration schemes for Hamiltonian systems with
  constraints such as SHAKE or RATTLE which preserve holonomic
  constraints encounters difficulties when applied to the numerical
  treatment of the equations of general relativity.
\end{abstract}
\maketitle

It is well known that the equations of General Relativity (GR) can be
derived from a variational principle and that they can be cast into
Hamiltonian form. The underlying symplectic structure has been studied
as early as the 1940's beginning with the work of
Bergmann~\cites{Bergmann:1949p2751,Anderson:1951p2747},
Dirac~\cites{Dirac:1958p1440,dirac2003:lectures-on-quantum-mechanics}
and ADM~\cite{arnowittdeser62}. The main motivation then has been to
work out a quantisation scheme for GR. For various reasons, not
the least of them being the peculiar nature of the symplectic
structure of GR, these early attempts have not led to any viable
theory of quantum gravity.

On the other hand it has been well established within the numerical
mathematics
community~\cites{Hairer:2003p187,Hairer2002:Geometric-numerical-integration,McLachlan:2006p178}
that the use of so called symplectic integrators i.e., numerical ODE
solvers which preserve an underlying symplectic structure can lead to
significant improvements in long-time stability, conservation of first
integrals and accuracy. These methods have been generalised even to
Hamiltonian systems with constraints. There are two particularly
noteworthy methods which are called SHAKE~\cite{Ryckaert:1977p184} and
RATTLE~\cite{Anderson:1982p183}. They have been developed within the
area of molecular dynamics but they have since then been used
successfully in various other applications. However, they only work
for holonomic constraints.

Given the success of these methods it is, therefore, natural to apply
symplectic numerical methods also to the equations of GR. However, as
we will argue in this paper, it is not clear (yet) whether there is any
advantage to be gained in this approach.

This paper addresses the question of the applicability of symplectic
integrators in GR and it is directed towards both communities,
numerical mathematics as well as numerical relativists. This
necessarily means that we need to review both the Hamiltonian
framework for GR as well as the essence of symplectic
integrators. This is reflected in the structure of the paper which
consists mostly of sections to introduce the necessary background
material. In sect.~\ref{sec:hamiltonian-systems} we describe
finite-dimensional Hamiltonian systems and in
sect.~\ref{sec:constr-hamilt-syst} we expand this to include systems
with constraints. Sect.~\ref{sec:sympl-struct-gr} is devoted to a
brief exposition of the symplectic structure of GR in the special case
of spatially compact space-times. In sect.~\ref{sec:sympl-integr} we
describe the essential properties of symplectic integrators for
constrained systems. Finally, in sect.~\ref{sec:conclusion}  we
discuss the consequences of trying to combine these two areas of
research.

\section{Hamiltonian systems}
\label{sec:hamiltonian-systems}

Before we come to the symplectic structure of GR let us first look at
a classical Hamiltonian system with finitely many degrees of freedom
such as those occurring in classical mechanics, molecular dynamics
etc. The system is specified by a triple~$(\cP,\omega,H)$, where $\cP$
is a real manifold of even dimension $2n$ which carries a
symplectic form $\omega$, i.e., a non-degenerate closed 2-form. The
pair $(\cP,\omega)$ is called the phase space of the system. It is the
collection of all states which are accessible to the system and the
symplectic form provides a way to locally sort the degrees of freedom
into pairs of conjugate variables.

Since $\omega$ is non-degenerate it defines at each point $x\in\cP$ an
isomorphism between the tangent space $T_x\cP$ and the co-tangent
space $T^*_x\cP$. Thus, any function $f \in \cC^\infty(\cP)$ defines a
Hamiltonian vector field $X_f$ by the equation
\begin{equation}
  \label{eq:iXomega}
  {X_f}  \hook \omega + \dd f = 0.
\end{equation}
From this equation and the closure of $\omega$ follows that the Lie
derivative 
\begin{equation}
  \label{eq:LXomega}
  \Lie_{X_f} \omega = \dd({X_f} \hook \omega) + {X_f} \hook \dd\omega = 0,
\end{equation}
i.e., the symplectic form is invariant under the flow generated by a
Hamiltonian vector field. Each member of the flow is a canonical
transformation. This is true, in particular, for the Hamiltonian
vector field $X_H$ generated by the Hamiltonian function $H:\cP \to
\RR$, the function which specifies the dynamics of the system; the
time evolution map $\phi_t$ generated by $X_H$ which maps an arbitrary
initial state $x_0\in\cP$ to the state at time $t$ is a canonical
transformation.

Dual to the symplectic form we can introduce a Poisson structure,
i.e., Poisson brackets
$\{\cdot,\cdot\}$ on $\cP$, by defining for any two functions $f,g \in
\cC^\infty(\cP)$ 
\begin{equation}
  \label{eq:poisson}
  \{f,g\} := \omega(X_f,X_g) = \Lie_{X_f} g.
\end{equation}
This turns the algebra of functions on $\cP$ into a Lie algebra with
respect to the Poisson bracket, the Jacobi identity being a
consequence of the closure of $\omega$. It is well known that there
exist preferred so called canonical coordinates $(p_k,q^i)$ on $\cP$
such that locally the symplectic form is
\[
\omega = \dd p_k \wedge \dd q^k
\]
or, equivalently, such that these coordinates have canonical
commutation relations
\[
\{p_i,p_k\} = 0, \qquad \{q^i, q^k\} = 0, \qquad \{p_i,q^k\} = \delta^k_i.
\]
The flow generated by a function $H\in \cC^\infty(\cP)$ induces a
change in a function $f$ which is given by
\[
\dot f = \{H,f\}.
\]
In particular, the rate of change in the canonical variables can be
used to obtain a coordinate expression for the flow
\[
\dot q^i = \{H,q^i\}, \qquad \dot p_k = \{H,p_k\}.
\]

Hamiltonian systems frequently arise from Lagrangian systems by
performing a Legendre transformation. The most common case is where
the Lagrangian system is defined by an action functional 
\[
\cA = \int \cL(q,\dot q)\, \dd t
\]
over a Lagrangian function $\cL: TQ \to \RR$ on the tangent bundle of
a configuration manifold~$Q$. A Legendre transformation is then used
to define a Hamiltonian system on the cotangent bundle $T^*Q$ of the
configuration space. A detailed description of these structures can be
found e.g., in
\cites{abrahmarsd1978:foundations-of-mechanics,woodh1995:geometric-quantization,arnol1978:mathematical-methods-of-classical}

In many cases the Legendre transformation is well-defined and
invertible and the Hamiltonian system is valid without any
restrictions, i.e., it is unconstrained. In some cases, however, when
the Lagrangian function is degenerate, the Legendre transformation is
not a local diffeomorphism. This implies that not all the possible
states in $T^*Q$ are available to the Hamiltonian system, i.e., that
there are constraints which have to be imposed.

This situation has been analysed in detail by
Dirac~\cites{Dirac:1958p1440,dirac2003:lectures-on-quantum-mechanics} (see
also~\cite{henneteite1992:quantization-of-gauge-systems}) who
developed a theory of Hamiltonian systems with constraints.

\section{Constraints in Hamiltonian systems}
\label{sec:constr-hamilt-syst}

From the geometric point of view a constraint in a
phase space~$(\cP,\omega)$ is a sub-manifold $\cC$ of $\cP$ which
comprises the states which are accessible to the system. The
symplectic form $\omega$ restricts to a closed 2-form $\bar\omega$ on
$\cC$. In general, $\bar\omega$ will not be regular. Let
\[
G_x=\{U \in T_x\cC :\bar\omega(U,V)=0, \,\forall V\in T_x\cC\}.
\] 
At each $x\in \cC$ this is a subspace of $T_x\cC$ and we assume that
the dimension of $G_x$ is constant as $x$ varies over $\cC$. Then
$G=\bigcup G_x$ is a sub-bundle of $T\cC$ (and hence also of $T\cP$)
which defines a distribution in $T\cC$. It is easily seen that this
distribution is integrable: let $X$, $Y$ be two sections of $G$, so
that $X\hook \bar\omega=Y \hook \bar\omega = 0$. Then
the closure of $\bar\omega$ implies
\[
[X,Y] \hook \bar\omega = \Lie_X (Y \hook \bar\omega) - Y \hook \Lie_X
\bar\omega = - Y \hook \dd (X \hook \bar\omega) - Y \hook (X \hook \dd
\bar \omega) = 0 .
\]
Therefore, there exist maximal integral surfaces $\cG$ tangent to $G$
which foliate $\cC$. Under certain technical assumptions (for the
details see~\cite{woodh1995:geometric-quantization} and references
therein) the space of leaves $\cP'=\cC\vert_\cG$ is a differentiable
manifold. Furthermore, there exists a closed 2-form $\omega'$ on
$\cP'$ which pulls back to $\bar\omega$ under the canonical projection
and which is regular. Thus, the pair $(\cP',\omega')$ is a phase space
on its own.

This is all that can be said from `inside~$\cC$', i.e., without taking
into account that $\cC$ is in fact a sub-manifold of $\cP$. Doing
this, one obtains information about how the embedding of~$\cC$
in~$\cP$ affects the structure inside~$\cC$. Let us first define
\[
T_x\cC^\perp = \{U \in T_x\cP :\omega(U,V)=0, \,\forall V\in T_x\cC\}.
\]
then, clearly, $G_x = T_x\cC\cap T_x\cC^\perp$. Let $r$ be the
co-dimension of~$\cC$ in $\cP$, then we have $\dim T_x\cC = 2n-r$ and
$\dim T_x\cC^\perp = r$. Furthermore, let $f \in \cC^\infty(\cP)$ be
constant on $\cC$, so that the restriction of $\dd f$ to $\cC$
vanishes. Then at all $x\in\cC$ we have for any $V \in T_x\cC$
\[
\omega_x(X_f,V) = -V(f) = 0
\] 
i.e., $X_f(x) \in T_x\cC^\perp$. Let $C_A$ be $r$ independent
functions which vanish on~$\cC$ near $x$ so that~$\cC$ may locally be
regarded as the zero-set of these functions. Clearly, any function
which is locally constant on~$\cC$ is functionally dependent on
the~$C_A$. Hence, the Hamiltonian vector fields $X_A:=X_{C_A}$
evaluated at~$x$ generate a $r$-dimensional vector space which,
therefore, coincides with~$T_x\cC^\perp$.

The vector fields $X_f$ for locally constant~$f$ need not be tangent
to~$\cC$. We will be interested mostly in two cases: when either none
or all of the vector fields are tangent to~$\cC$.

In the first case we have $G_x = T_x\cC \cap T_x\cC^\perp = \{0\}$, so
that $\bar\omega$ is regular and $\cP' = \cC$. Then $(\cC,\bar\omega)$
is a symplectic sub-manifold of $(\cP,\omega)$ i.e., it is a
phase space in its own right. Note, that $Q_{AB}:= \omega(X_A,X_B) =
\{C_A,C_B\}$ is a non-singular $r\times r$-matrix when evaluated
on~$\cC$. In this case, the constraint functions $C_A$ are called
\emph{second class constraints}.

Since $(\cC,\bar\omega)$ is a phase space there exists also a Poisson
bracket on $\cC$ corresponding to $\bar\omega$ defined for functions
on~$\cC$.  Denoting the inverse of $Q_{AB}$ by $Q^{AB}$, so that
$Q_{AB} Q^{BC}=\delta_A^C$ we can express the Poisson bracket $\{\bar
f, \bar g\}$ between two functions $\bar f$ and $\bar g$ on~$\cC$ in
terms of Poisson brackets on~$\cP$ as follows. Choose extensions of
$\bar f$ and $\bar g$ to $\cP$, i.e., functions $f$ and $g$ on $\cP$
which restrict to $\bar f$ and $\bar g$ on $\cC$. Then, on $\cC$ the
following equation holds:
\begin{equation}
  \label{eq:diracbracket}
  \{\bar f,\bar g\} = \{f,g\} - \{f,C_A\}Q^{AB}\{C_B,g\} .  
\end{equation}
Here, the left hand side is the Poisson bracket on $(\cC,\bar\omega)$
and it is defined only on $\cC$ while the right hand side is
well-defined even on $\cP$. It makes sense for arbitrary functions $f$
and $g$. It is easy to see that it vanishes if $f$ or $g$ are taken as
constraints. Since two extensions of $\bar f$ coincide on $\cC$ they
differ by constraints. This shows that it is irrelevant which
extensions for $\bar f$ or $\bar g$ are used. The expression on the
right hand side satisfies the defining properties of a Poisson
structure so we may also regard it as defining a new Poisson bracket
$\{\cdot,\cdot\}_D$ on~$\cP$, which is adapted to the existence of the
constraint surface. This new Poisson bracket is called Dirac
bracket~\cite{dirac2003:lectures-on-quantum-mechanics}. Note, that we
can now express the Poisson bracket on~$\cC$ in terms of Dirac's
bracket
\[
\{\bar f, \bar g\} = \{f,g\}_D,
\]
which in turn enables us to discuss the Poisson structure of
constrained system in terms of quantities on the original phase space.

The second case of interest is characterised by the fact that
all the Hamiltonian vector fields $X_A$ corresponding to constraint
functions are tangent to $\cC$. Therefore, we have $T_x\cC^\perp
\subset T_x\cC$ and $G_x=T_x\cC^\perp$. This implies, that
\[
\{C_A, C_B\} = \omega(X_A,X_B) = \Lie_{X_A} C_B 
\]
which vanishes on $\cC$. It has been useful to introduce the notion of
`weak equality' of two functions $f$ and $g$ if and only if they
restrict to the same function on~$\cC$. Thus,
\[
f \approx g \iff f-g = \mu^A C_A
\]
for appropriate functions $\mu^A \in \cC^\infty(\cP)$. Hence, in the
present case we may write
\[
\{C_A,C_B\} \approx 0.
\]
In this case, the functions $C_A$ which define the constraint
hypersurface are in involution. They are called \emph{first class
constraints}.

Since $G_x \ne \{0\}$ the restriction of the symplectic
form~$\bar\omega$ is degenerate and $(\cC,\bar\omega)$ is a
pre-symplectic manifold. Factoring out the leaves of the foliation we
obtain the reduced phase space $(\cP',\omega')$, sometimes called the
space of the true degrees of freedom.

Let us now consider time evolution. Given a Hamiltonian $H \in
\cC^\infty(\cP)$ for a system with constraints we need to ask for
compatibility of the time evolution generated by $H$ with the
constraints: when the system is started out on~$\cC$ then it should
remain on~$\cC$ i.e., the Hamiltonian vector field $X_H$ should be
tangent to~$\cC$ or, expressed in terms of Poisson brackets, the weak
equality
\begin{equation}
\{ H, C_A\} \approx 0\label{eq:compatibility}
\end{equation}
should hold for all constraints~$C_A$. Clearly, for the behaviour of
the constrained system only the restriction $\bar H$ of the
Hamiltonian function to~$\cC$ is relevant and the extensions of $\bar
H$ to $\cP$ (of which $H$ is one) are all a priori
equivalent. However, we may try to find a compatible extension $\tilde
H$ for which the Hamiltonian vector field~$X_{\tilde H}$ is tangent
to~$\cC$. Writing $\tilde H=H+\lambda^B C_B$ we find
\[
0 \approx \{\tilde H,C_A\} = \{H,C_A\} + \lambda^B\{C_B,C_A\} +
\{\lambda^B,C_A\} C_B \approx \{H,C_A\} + \lambda^B Q_{BA} .
\]
This equation tells us that we can find a compatible extension only if
$Q_{AB}$ is invertible, i.e., only if the constraints are second
class. Only in this case we can express the dynamics of the constrained
system entirely in terms of the original phase space $\cP$.

In the case of first class constraints $Q_{AB}\approx 0$ so that
either all extensions or none satisfy the compatibility
condition~\eqref{eq:compatibility}. If it is satisfied then $H$ is
constant along the Hamiltonian vector fields $X_A$ generated by the
constraints $C_A$. Hence, it descends to a well-defined function on
$\cP'$. Furthermore, for its Hamiltonian vector field $X_H$ we have
\[
[X_H,X_A] \hook \omega = - \dd(\{H,C_A\}) .
\]
Since for any weakly vanishing function $f\approx0$ one has $\dd f =
\dd \lambda^AC_A + \lambda^A \dd C_A$ for suitable functions
$\lambda^A$ this implies that for any $x\in\cC$ and $Y\in T_x\cC$ 
\[
\omega([X_A,X_H],Y) = \bar\omega([X_A,X_H],Y) = \lambda^B Y(C_B) = 0.
\]
Thus, $[X_A,X_H] \in T_x\cC^\perp$ so that 
\[
\Lie_{X_A} X_H \in G_x.
\]
This implies that $X_H$ is projectable onto $\cP'$. One can also
easily see, that its projection is the Hamiltonian vector field for
the projected Hamiltonian with respect to the symplectic form
$\omega'$.

Let us now illustrate the two cases with two examples.

\subsection{Example 1: a particle restricted to a hypersurface}
\label{sec:exampl-part-restr}

Consider a free particle in a Riemannian manifold $(Q,g_{ab})$ whose
motion is restricted to a hypersurface $S\subset Q$. Let $C_0 = F$ be
a function whose zero-set locally defines~$S$. In local coordinates
$q^a$ on $Q$ the action for this situation is given by
\[
\cA = \int \left(\frac12 m g_{ab}(q) \dot q^a \dot q^b - \lambda F(q)
\right)\,\dd t
\]
This leads to the Hamiltonian $H=\frac1{2m}g^{ab} p_a p_b + \lambda
F(q)$. Requiring that $\{H,C_0\} \approx 0$ gives us (using the notation
$F_a = \nabla_a F$)
\[
C_1 :=p^a F_a \approx 0
\]
so we need to include $C_1$ as a constraint. Since $\{C_0,C_1\} =
-F_aF_bg^{ab} \ne 0$ we can solve the equations
\[
\{H+\lambda_0 C_0 + \lambda_1 C_1,C_i\} \approx 0 \text{ for } i=0,1
\]
for $\lambda_0$ and $\lambda_1$ and obtain
\[
\lambda_1 = 0, \qquad \lambda_0 = \frac1m \frac{p^ap^b F_{ab}}{p^ap_a}
\]
with $F_{ab}=\nabla_a\nabla_bF$.
Hence, the final Hamiltonian is
\[
H = \frac1{2m} p^a p_a -\frac1m \frac{p^ap^b F_{ab}}{p^ap_a}\,C_0.
\]
It is straightforward to check that its Hamiltonian vector field annihilates
both constraints.

\subsection{Example 2: relativistic particle}
\label{sec:exampl-relat-part}

We consider a particle in a Lorentzian space-time $(Q,g_{ab})$. In
this case the action for the world-line $q^a(\tau) \in Q$ of the
particle is given by
\[
\cA = m \int \sqrt{g_{ab}\dot q^a \dot q^b}\, \dd \tau.
\]
The distinguishing feature of this action is its invariance under
reparametrisation, $\tau \mapsto \tau'=T(\tau)$. The conjugate
momentum is
\[
p_a = \frac{m}{\ell} \dot q^b g_{ab},
\]
where we abbreviate $\ell=\sqrt{g_{ab}\dot q^a \dot q^b}$. Obviously,
we obtain the relation
\begin{equation}
C(p,q): = g^{ab}p_ap_b - m^2 = 0,\label{eq:constraint}
\end{equation}
i.e., the momenta cannot attain all possible values. Hence, the states
of the system are confined to the sub-manifold $\cC\subset T^*Q$
defined by~(\ref{eq:constraint}). From this constraint we obtain the
further relation
\begin{equation}
p^a\dd p_a = 0\label{eq:pdp}
\end{equation}
which holds on~$\cC$. The restriction of $\omega$ to $\cC$ has a
kernel which we can determine as follows. Let $X=X^a\del/\del q^a +
Y_b\del/\del p_b$ then we search for non-vanishing $X$ on $\cC$ with
\[
0 = X\hook \omega = Y_b\dd q^b - X^a\dd p_a
\]
which, in view of~\eqref{eq:pdp} implies $Y_b=0$ and $X^a=\alpha
p^a$ for an arbitrary function $\alpha$ on $\cC$. Thus, every vector
field in the kernel of $\bar\omega$ has the 
form
\[
X= \alpha p^a\frac{\del}{\del q^a}.
\]
Since the kernel is 1-dimensional the vector fields are proportional
to each other and their integral curves coincide as sets. It is not
difficult to show that these vector fields generate exactly the
reparametrisation along the integral curves, i.e., they generate
gauge-transformations.

The Hamiltonian vector
field of the constraint~$C$ is also in the kernel of $\bar\omega$
\[
X_C = 2 p^a \frac{\del}{\del q^a} 
\]
so that it is tangent to $\cC$. It generates the flow
\[
\phi_\lambda(p_a,q^a) = (p_a, q^a+2\lambda p^a).
\]
The Hamiltonian function can be determined from the Lagrangian in the
usual way
\[
H = p_a\dot q^a - m \ell = \frac{m}\ell g_{ab}\dot q^a \dot q^b - m
\ell = 0. 
\]
Clearly, this Hamiltonian is compatible with the constraints. In fact,
it vanishes on $\cC$ which is consistent with the fact that it
generates gauge transformations.

Thus, we have the following picture. The system does not specify
individual points $(p_a,q^a)\in \cC$ as its states but instead one
should regard as one state the collection of all points which lie on
the same integral curve of the gauge vector fields $X$. They must be
considered as equivalent because they are related by some
gauge-transformation. Hence, the states of the system are global
entities, an entire world-line considered as a point set i.e., without
a distinguished parametrisation.

Since the Hamiltonian vanishes on $\cC$ it is functionally dependent
on the constraint and it also generates a gauge-transformation. So in
this sense there is no distinguished time evolution in this system
which would map from one state to another as it is the case in many
`normal' systems.

If one is interested in the structure of an individual world-line then
one can proceed by fixing an initial point on the line and then,
using the Hamiltonian vector field of $H$, the integral curve through
that point can be found. However, the result will be a curve together
with a special parameter which is determined by the choice of the
Hamiltonian.  The system of a relativistic particle is very similar to
the situation in GR to which we will now turn.

\section{The symplectic structure of GR}
\label{sec:sympl-struct-gr}

We now come to a brief introduction to the symplectic structure of
GR. We follow loosely the exposition in~\cite{ashtekar88:_new}. Other
treatments can be found in
e.g.,~\cites{woodh1995:geometric-quantization,%
frauendiener1992:_sympl_form,%
wald84:_gener_relat,%
ashtekarbombelli91:_covar_phase_space}.
Let $\Sigma$ be a 3-dimensional compact closed manifold\footnote{We
  concentrate here on the case of spatially closed space-times because
  we are interested in the intrinsic Hamiltonian framework. Issues
  concerning boundary conditions like in the case of asymptotically
  flat space-times or even in the quasi-local regime are somewhat
  cumbersome to formulate or are not even resolved
  yet~\cite{Szabados:2004p218}.}. We consider globally hyperbolic
space-times of the form $M=\Sigma\times \RR$. We choose a global
time-function $t:\Sigma \times \RR \to \RR$ and a vector field $t^a$
such that the hypersurfaces $\Sigma_t$ of constant $t$ are
diffeomorphic to $\Sigma$ and such that $t^a\del_a t =1$. We assume
that the hypersurfaces $\Sigma_t$ are space-like and that the vector
field $t^a$ is future directed and time-like. Let $n_a$ be the future
directed co-normal of the hypersurfaces and denote by ${}^4g_{ab}$
resp. $g_{ab}$ the space-time metric resp. the metric on $\Sigma_t$.

We can perform a $3+1$-decomposition of the geometrical quantities in
the usual way~\cite{wald84:_gener_relat} by writing $t^a = \alpha n^a +
\beta^a$, thereby introducing the lapse function $\alpha$ and the shift vector
$\beta^a$. Thus, we can express the 4-geometry in terms of (families of)
3-dimensional quantities. In this way the Einstein-Hilbert action
\begin{equation}
  \label{eq:einsteinhilbert}
  \int_{\Sigma\times\RR} {}^4R\,\sqrt{-{}^4g}\; \dd^4x
\end{equation}
can be expressed up to boundary terms as the following action
\begin{equation}
  \label{eq:3action}
  \cA = \int_\RR  \cL(g,\dot g;\alpha,\beta) \, \dd t
\end{equation}
where the Lagrangian is
\begin{equation}
  \label{eq:lagrangean}
  \cL(g,\dot g;\alpha,\beta) = \int_\Sigma \alpha\left( R +
    K^{ab}K_{ab} - K^2 \right) \sqrt{g}\, \dd^3x .
\end{equation}
Here, we have used the scalar curvature $R$ of the metric $g_{ab}$ on
$\Sigma_t$, the extrinsic curvature $K_{ab}$ and its trace $K=K_c{}^c$
of $\Sigma_t$ within the space-time $M$. Due to the relationship 
\[
2 \alpha K_{ab} = \dot g_{ab} - (\Lie_\beta g)_{ab}
\] 
between the extrinsic curvature and the Lie derivative $\dot g_{ab}:=
(\Lie_t g)_{ab}$ of the metric the Lagrangian  is considered as a
functional of $g_{ab}$, its time derivative $\dot g_{ab}$ as well as
the lapse and shift. Note, that $\cL$ does not contain any time
derivatives of $\alpha$ or $\beta^a$ which indicates that it is
singular. In fact, computing the variations of $\cL$ with respect to
$\alpha$ and $\beta^a$ yields
\begin{equation}
  \label{eq:constraints}
  C \equiv  \frac{\delta\cL}{\delta \alpha} = \sqrt{g} \left(R - K^{ab}K_{ab} +
    K^2 \right),\quad
  C_a \equiv \frac{\delta\cL}{\delta \beta^a} = 2 \sqrt{g} \,\nabla_b
  \left(K^b{}_a - \delta_a^b\,K \right).
\end{equation}
The vanishing of these expressions as required by the Euler-Lagrange
equations yields constraints on the possible configurations.

In a similar way we compute the momentum conjugate to $g_{ab}$ as
\begin{equation}
  \label{eq:conjmomentum}
  p^{ab} \equiv \frac{\delta \cL}{\delta \dot g_{ab}} = \sqrt{g} \left( K^{ab}
  - K g^{ab} \right).
\end{equation}
Note, that this and the constraint expressions are tensor valued
densities of weight~1.

Finally, we determine the Hamiltonian from the formula
\begin{equation}
  H(g,p) = \int_\Sigma \dot g_{ab} p^{ab}\,\dd^3x - \cL
\end{equation}
and find (up to boundary terms)
\begin{equation}
  \label{eq:hamiltonian}
  H(g,p) = \int_\Sigma \alpha \sqrt{g}\left[ - R +
    \frac1{g}(p^{ab} p_{ab}- \frac12 p^2)  \right] + \beta^b
  \left[ -2 \nabla_ap^a{}_b\right] \dd^3x. 
\end{equation}

Thus, we have the following situation. As the configuration space
$Q$ we take the space of Riemannian metrics on $\Sigma$. The tangent
space $T_{g}Q$ consists of all symmetric covariant second rank
tensor fields $\delta g_{ab}$ on $\Sigma$. The cotangent space
$T^*_{g}Q$ is defined as the space of functionally differentiable
1-forms on $T_g Q$, i.e., linear real-valued maps which are of the
form
\[
T_gQ \supset \delta g_{ab} \mapsto \int_\Sigma p^{ab} \delta g_{ab} 
\]
where $p^{ab}$ is a tensor valued density of weight~1. The phase space
$\cP$ of general relativity (in the context of spatially closed
space-times) is the cotangent bundle $T^*Q$ over the space $Q$ of
Riemannian metrics over $\Sigma$. Points of $\cP$ are represented as
pairs $(g_{ab},p^{ab})$ and tangent vectors to $\cP$ are represented
as pairs $(\delta g_{ab},\delta p^{ab})$. Being a cotangent bundle
$\cP$ carries a canonical symplectic form and hence also a Poisson
structure.

The symplectic form between two tangent vectors to $\cP$ is defined by
\begin{equation}
  \label{eq:sympformGR}
  \omega_{(g,p)}((\delta_1 g,\delta_1 p),(\delta_2 g,\delta_2 p)) =
  \int_{\Sigma} \delta_1p^{ab}\delta_2g_{ab} -
  \delta_2p^{ab}\delta_1g_{ab}\,\dd^3x
\end{equation}
and the corresponding Poisson bracket between two functions $F$ and
$G$ on $\cP$ is
\begin{equation}
  \label{eq:poissonGR}
  \left\{F,G\right\} = \int_\Sigma \frac{\delta F}{\delta
    p^{ab}}\frac{\delta G}{\delta g_{ab}} - \frac{\delta F}{\delta
    g_{ab}}\frac{\delta G}{\delta p^{ab}}\, \dd^3 x.
\end{equation}
The constraints expressions~(\ref{eq:constraints}) yield functions on
$\cP$ by integration over~$\Sigma$
\[
C_f = \int_\Sigma f C\,\dd^3x,\qquad C_{\mathbf{v}} = \int_\Sigma
v^a C_a\, \dd^3x,
\]
where $f$ and $\mathbf{v}=v^a$ are arbitrary test (vector) fields on
$\Sigma$. Using the Poisson bracket we can easily see that the
constraint functions satisfy the Poisson commutation relations
\begin{equation}
  \label{eq:poissonconstraints}
  \begin{aligned}
    \{C_f,C_g \} &= - C_{f\boldsymbol{\nabla}g - g\boldsymbol{\nabla}f},\\
    \{C_f,C_{\mathbf{v}} \} &= - C_{\mathbf{v}(f)},\\
    \{C_{\mathbf{v}},C_{\mathbf{w}} \} &= - C_{[\mathbf{v},\mathbf{w}]}.
  \end{aligned}
\end{equation}
Therefore, the Poisson brackets among all constraints are again
constraints, i.e., \emph{the constraints are first class}. The
constraint functions $C_f$ and $C_{\mathbf{v}}$ generate
transformations on $\cC$ which correspond to gauge-transformations,
thus mapping a state $(g_{ab},p^{ab})$ to an `equivalent' state. The
constraints $C_{\mathbf{v}}$ generate 3-dimensional diffeomorphisms
within $\Sigma$. The constraints $C_f$, however, generate
transformations between different hypersurfaces $\Sigma_t$ which can
be interpreted as the `evolution' of the intrinsic and extrinsic
geometry of $\Sigma$ within the space-time $M$ along the vector field
$t^a = f n^a$.

The Hamiltonian~(\ref{eq:hamiltonian}) turns out to be a combination
of constraints
\begin{equation}
  \label{eq:hamconstr}
    H(g,\pi) = C_\alpha + C_{\boldsymbol{\beta}}.
\end{equation}
Hence, it generates gauge-transformations, namely the evolution of
$\Sigma$ along the general evolution vector $t^a = \alpha n^a +
\beta^a$.  This implies that we have a similar situation here as in
the case of the relativistic particle. A particular given state
$(g_{ab},p^{ab})$ on~$\cC$ is equivalent to states $(\hat g_{ab}, \hat
p^{ab})$ which are obtained by such transformations. Each equivalence
class corresponds to the same single space-time. 

The fact that GR is a completely constrained system is the Hamiltonian
way of reinstating general covariance of the theory. Any
time-evolution in the Hamiltonian sense would map equivalence classes
to equivalence classes. i.e., a space-time to an entirely different
space-time which would not make any sense. Instead the Hamiltonian
formulation of GR specifies the general covariant geometry of a single
space-time eliminating any allusion to a notion of time.

\section{Symplectic integrators}
\label{sec:sympl-integr}

Let $(\cP,\omega,H)$ be a (finite-dimensional) Hamiltonian system
possibly with constraints. The flow generated by $H$ maps initial
states $x_0$ to later states $x_t=\phi_t(x_0)$. The map $\phi_t:\cP
\to \cP$ is a canonical map, the `time-$t$' map. It is obtained by
finding the integral curves of the Hamiltonian vector field of $H$,
i.e., by solving a system of ODE when expressed in canonical
coordinates.

There are many methods to solve systems of ODE by numerical
means. Some of them have the special property that they preserve the
structure defining the Hamiltonian system. We may regard a numerical
method as a map $\Phi_h:\cP \to \cP$ which maps a state $x_n$ to the
next state $x_{n+1}$ and we call such a method a symplectic integrator
(of order $p$) if $\Phi_h$ is a canonical transformation for every $h$
which approximates the exact Hamiltonian flow for a Hamiltonian
function $H$ in the sense that
\begin{equation}
  \label{eq:symplintp}
  \Phi_h(x) = \phi_h(x) + \cO(h^{p+1})
\end{equation}
for all $x\in \cP$. In~\cite{Hairer:1994p2702} it is shown that a
symplectic integrator of order $p$ is backward stable i.e., that there
exists a Hamiltonian function $\tilde H_h$ such that $\tilde H_h - H =
\cO(h^{p+1})$ and such that $\Phi_h$ is the time-$h$ map of the
Hamiltonian vector field corresponding to $\tilde H_h$. This means
that a symplectic method can be regarded as the exact time-$h$ map for
a slightly perturbed Hamiltonian system.

When constraints are present the symplectic integrators can be
generalised to numerical methods which preserve the symplectic
structure and the constraint hypersurface simultaneously~\cites{Hairer2002:Geometric-numerical-integration,Reich:1996p195}. Examples of
such methods are the well-known algorithms
SHAKE~\cite{Ryckaert:1977p184} and RATTLE~\cite{Anderson:1982p183}
developed within the context of molecular dynamics. They are
implemented schematically as follows. Consider the Hamiltonian system
$(\cP,\omega,H)$ together with constraints $C_A$ and let $x_0=(p_0,q_0)$
be a point on the constraint hypersurface $\cC$. We seek a method to
compute the next point $x_h=(p_h,q_h)$ after time $h$ on $\cC$ according
to the Hamiltonian $H$. One considers the extended Hamiltonian
\[
\bar H = H + \lambda^A C_A
\]
which generates the equations of motion on $\cC$
\begin{equation}
  \label{eq:eqm}
  \dot p = - \frac{\del H}{\del q} - \lambda^A \frac{\del C_A}{\del q},\quad
  \dot q = \frac{\del H}{\del q} + \lambda^A \frac{\del C_A}{\del p}.
\end{equation}
These have the approximate solutions
\begin{equation}
  \label{eq:approximatesol}
  \begin{aligned}
    p_h = \left(p_0 - h \frac{\del H}{\del q}(x_0)\right) -
    h \lambda^A \frac{\del C_A}{\del q}(x_0) + \cO(h^2),\\
    q_h = \left(q_0 + h \frac{\del H}{\del p}(x_0)\right) + h
    \lambda^A \frac{\del C_A}{\del p}(x_0) + \cO(h^2).
  \end{aligned}
\end{equation}
However, the multipliers $\lambda_A$ are not yet known. They are
determined by requiring that the point $x_h=(p_h,q_h)$ lies on
$\cC$. Thus, one puts
\[
\hat p = p_0 - h \frac{\del H}{\del q}(x_0),\quad
\hat q = q_0 + h \frac{\del H}{\del p}(x_0)
\]
and notes that
\[
\begin{aligned}
  C_B(x_h) &= C_B(\hat x) - h \frac{\del C_B}{\del p}(\hat x)
  \frac{\del C_A}{\del q}(x_0) \lambda^A + \frac{\del C_B}{\del
    q}(\hat x) \frac{\del C_A}{\del
    p}(x_0) \lambda^A + \cO(h^2)\\
  &= C_B(\hat x) - h \lambda^A \left\{ C_B, C_A\right\}(\hat x) +
  \cO(h^2).
\end{aligned}
\]
Thus, one can find the multipliers $\lambda^A$ by iteratively solving
the linear equation
\begin{equation}
C_B(\hat x) - h  \lambda^A \left\{ C_B, C_A\right\}(\hat x) =
0.\label{eq:lineareq}
\end{equation}
At each step the $\lambda^A$ are used to update $\hat x$, thus
entering a new iteration until the constraints $C_B(\hat x) = 0$ are
satisfied to a desired accuracy at which point one puts $x_h=\hat x$.

Due to the special structure of holonomic constraints and their
associated `hidden' constraints the SHAKE and RATTLE algorithms differ
in the details of this iteration procedure but the general structure
of the algorithms is as indicated here. The main point about them is
that they make the tacit assumption that \emph{the matrix $Q_{AB}
  = \{C_A,C_B\}$ is invertible at every~$\hat x$}. This implies that
these algorithms \emph{work only for second class constraints}. In
fact, the above calculation is nothing but a variant of the
calculation to find an extension of $H|_\cC$ whose Hamiltonian vector
field is tangent to $\cC$.

\section{Conclusion}
\label{sec:conclusion}

We have seen in sect.~\ref{sec:sympl-struct-gr} that GR is a fully
constrained theory with first class constraints. All the
Hamiltonians~(\ref{eq:hamiltonian}) are combinations of
constraints generating gauge-transformations. So, strictly speaking,
there is no time-evolution. However, within computational gravity one
uses numerical methods to compute the geometry and hence the physics
of one particular space-time. In terms of the Hamiltonian framework
this can be understood as follows. 

Fix initial data, i.e., a point $(g_{ab},p^{ab})$ on the constraint
surface~$\cC$ and specify a particular Hamiltonian by fixing lapse
function and shift vector. This Hamiltonian generates a gauge flow
which maps the initial point to points which correspond to
hypersurfaces at a `later' coordinate time. This `evolution' is
clearly symplectic and it preserves the constraints. Hence, one can
try to use symplectic integrators for the task of determining the
geometry of the space-time in a particular gauge.

Suppose that we have arranged a spatial discretisation of the infinite
dimensional system which results in a finite dimensional Hamiltonian
system. This means that the discretisation results in a system of ODE
which is Hamiltonian with respect to the discretised symplectic form
and which preserves the discretised constraints. This can be achieved
by an
appropriate discretization of the action and then performing a
Legendre transformation\footnote{It is an interesting and open
  question as to how much structure of the continuous Hamiltonian
  system can be carried over to the discrete system.}.  Let $\Delta$ be a
parameter which measures the discretisation error. The discretisation
should be consistent with the continuous system in the sense that we
recover the latter from the former in the limit $\Delta\to0$.

Following the implementation of a symplectic integrator we determine
the equations of motion from an extended Hamiltonian
$H+\lambda^AC_A$. Note, that the index $A$ ranges over four times the
number of degrees of freedom used in the discretisation. As
demonstrated in sect.~\ref{sec:sympl-integr} the method relies on the
invertibility of the matrix $Q_{AB}=\{C_A,C_B\}$.

Now two things may happen. Either the Poisson brackets of discretised
constraints vanish on the constraint surface i.e., they are also first
class with respect to the discretised symplectic structure. Then the
matrix $Q_{AB}$ is not invertible and the symplectic integrator
algorithm fails. 

The other possibility is that the Poisson brackets of the discretised
constraints do not vanish which means that the matrix $Q_{AB}$ could
be invertible so that multipliers $\lambda^A$ could be found. However,
consistency requires that in the limit of vanishing $\Delta$ one
recovers the continuous system from the discrete one. And this in turn
implies that in that limit the conditioning of the matrix $Q_{AB}$
will become increasingly bad so that the linear
equation~(\ref{eq:lineareq}) cannot be reliably solved
anymore. Therefore, the continuum limit $\Delta\to0$ will result in
increasingly inaccurate discrete approximations to the real solution
in contrast to expectations. 

These consequences are observed in numerical implementations of the
Einstein equations which make use of symplectic integration
techniques~\cite{RichterLubich}.

The question of how to treat Hamiltonian systems with first class
constraints numerically appears to be an open issue within the theory
of symplectic integrators. At the moment there is no straightforward
remedy to these shortcomings. One possibility to circumvent the
consequences could be to change the system. Recall that we have chosen
a Hamiltonian by fixing lapse function and shift vector arbitrarily
but independently of the evolution. One way to proceed might be to
couple the choice of these gauge functions to the Hamiltonian
system. This could break the general covariance in such a way that the
resulting system has only second class constraints. However, exactly
how to proceed remains largely unclear (see~\cite{RichterLubich2} for
a recent approach).

Another issue of relevance here is the relationship between holonomic
constraints with their hidden constraints on the one hand and the
first class/second class classification of constraints. Is it possible
to find gauge conditions i.e., a $3+1$ split and a choice of spatial
coordinates, which give second class constraints and can be regarded as
holonomic constraints in an appropriate generalised sense? These
issues need further clarifications.




\end{document}